\def   \ni {\noindent}

\def   \ssk {\vskip  5truept}

\def   \bsk {\vskip 15truept}

\def   \newline {\hfil\break}

\def   \vd   {\vec{d}}
\def   \Teta {\vec{\Theta}}
\def   \Npix {N_{pix}}
\def   \mC   {{\bf C}}
\def   \mT   {{\bf T}}
\def   \mN   {{\bf N}}
\def   \sig  {\sigma}

\documentstyle[epsfig]{article}
\begin{document}

\hsize 5truein
\vsize 8truein
\font\abstract=cmr8
\font\keywords=cmr8
\font\caption=cmr8
\font\references=cmr8
\font\text=cmr10
\font\affiliation=cmssi10
\font\author=cmss10
\font\mc=cmss8
\font\title=cmssbx10 scaled\magstep2
\font\alcit=cmti7 scaled\magstephalf
\font\alcin=cmr6 
\font\ita=cmti8
\font\mma=cmr8
\def\ref{\par\noindent\hangindent 15pt}
\null


\title{\ni APPROXIMATING THE LIKELIHOOD FUNCTION OF CMB EXPERIMENTS}                                               

\bsk \bsk
\author{\ni J.G.~Bartlett, A.~Blanchard, M.~Douspis \& M.~Le~Dour}                                                       
\bsk
\affiliation{Observatoire de Strasbourg, 11 rue de l'Universit\'e,
	67000 Strasbourg, FRANCE
}                                                
\bsk
\baselineskip = 12pt

\abstract{ABSTRACT \ni
We discuss the problem of constraining cosmological parameters
with cosmic microwave background band--power estimates.  Because
these latter are variances, they do not have gaussian distribution
functions and, hence, the standard $\chi^2$--approach is not
strictly applicable.  A general purpose approximation to 
experimental band--power likelihood functions is proposed,  
which requires only limited experimental details.  Comparison
with the full likelihood function calculated for several 
experiments shows that the approximation works well.
}                                                    
\bsk
\baselineskip = 12pt
\keywords{\ni KEYWORDS:
}               

\bsk
\baselineskip = 12pt


\text{\ni 1. INTRODUCTION
\ssk
\ni

	Current cosmic microwave background (CMB) observations 
are already capable of constraining cosmological parameters 
(Bond \& Jaffe 1996; Lineweaver et al. 1997; Hancock et al. 1998;
Bartlett et al. 1998a).  The COBE data alone have 
for some time now been used to determine the amplitude 
and slope of the power spectrum of density perturbations, 
but newer data at smaller angular scales are now reaching
the so--called ``Doppler peaks''.
This is the key region which will decide the
fate of models such as inflation, cosmic defects or 
other contenters; this is the region which will, within
the context a particular model, impose strong 
constraints on the fundamental cosmological parameters,
such as the density parameter, $\Omega$. 

\begin{figure}
\centerline{\psfig{file=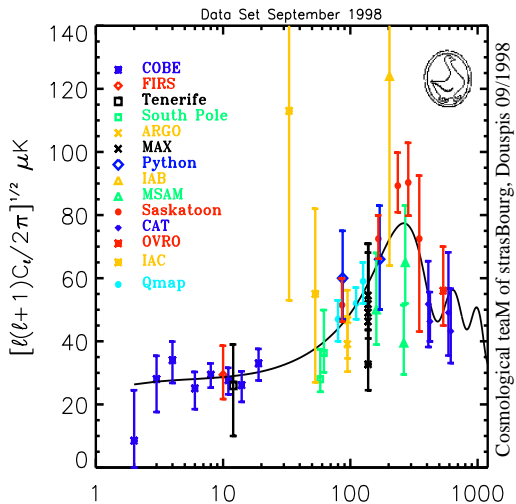, height=8cm,width=9cm}}
\caption{FIGURE 1. Current CMB data set, with the 
inflationary model prediction for $\Omega=0.9$,
$h=0.35$, $\Omega_bh^2=0.015$, $Q=17\mu$K and $n=0.94$. 
}
\end{figure}


	The present observational situation is summarized 
in Figure 1 as a set of band--power estimates over multipole
order $l$, the most common method of reporting 
CMB results.  Our discussion here will focus on statistical 
methods of constraining parameters within the context
of inflation--like models, which predict 
{\em gaussian} temperature fluctuations.  It is 
important to remark at this stage that the construction 
of Figure 1 {\em already supposes} gaussianity,
because the given band--power estimates assume gaussian
fluctuations (note that this includes the noise).  
This will become more clear in the following.
If the sky fluctuations where ever shown to be non--gaussian
(or an experiment contained an important component of
non--gaussian noise),
then the band--power estimates would need to be recalculated,
potentially changing Figure 1.  

	A common approach to parameter estimation 
applies the $\chi^2$--statistic to the band--power estimates
of Figure 1.  This technique, however, is not 
strictly applicable in this context because 
the points in Figure 1 are not gaussian distributed. 
This is true even if the underlying sky 
fluctuations, the pixel values (including noise), are in fact 
gaussian random variables.  Power estimates represent
the {\em variance} of the temperature
fluctuations, and an estimate of the variance of 
gaussian variables is not itself gaussian.
The $\chi^2$--statistic, which assumes
gaussianity, is therefore 
not the correct approach.  

	A rigorous analysis of the
data in Figure 1 requires that
the correct likelihood function be calculated
for each experiment.  Even with the present data
set, this is a time consuming task,
due to the variety of different 
experimental set--ups represented,
and for the next generation experiments,
with tens of thousands of pixels, it 
becomes computationally 
demanding in the extreme; for example,
analysis of BOOMERANG's North American
flight (30,000 pixels) requires $\sim 10$
hours on a Cray T3E (Borril 1998).
It is therefore important to 
find useful approximations to
experimental likelihood functions
(Bond et al. 1998; Wandelt et al. 1998).  
Here, we describe our efforts to derive a 
reasonable approximation.  Some preliminary 
results from its application to current data
can be found in Bartlett et al. (1998b).

\bsk
\ni
2 APPROXIMATE LIKELIHOOD FUNCTION
\ssk
\ni

	For gaussian anisotropies (gaussian noise is
assumed throughout), we
may write the likelihood function for a set of 
parameters, represented by a vector $\Teta$, 
once given the data, a set of pixel values 
arranged in a vector $\vd$:
\begin{equation}
\label{likeF}
{\cal L}(\Teta) \equiv {\rm Prob}(\vd|\Teta)
	= \frac{1}{(2\pi)^{\Npix/2} |\mC|^{1/2}} e^{-\frac{1}{2} 
	{\vd}^t\cdot \mC^{-1} \cdot \vd}
\end{equation} 
where $\mC$ is the covariance matrix
\begin{equation}
C_{ij} = <d_id_j>_{ens} = T_{ij} + N_{ij}
\end{equation}
The average is understood to be over the theoretical
ensemble of all possible anisotropy patterns
realizable with the given parameter set $\Teta$,
of which the actual data set is but one
realization.  The covariance has two contributions,
one intrinsic to the sky fluctuations, $\mT(\Teta)$, a 
function of the parameter vector, and the
other due to the noise, $\mN$.  For a data vector
consisting of simple pixel values, from a 
map of the sky, we further have
\begin{equation}
\label{eq:Tij}
T_{ij} = \frac{1}{4\pi} \sum_l (2l+1) C_l |B_l|^2 P_l(\cos\theta_{ij})
\end{equation}
where, as usual, the power spectrum is the ensemble 
of $C_l$, $B_l$ describes the experimental beam
(assumed spherically symmetric), $P_l$ is the Legendre 
polynomial of order $l$ and $\theta_{ij}$ is the
angle separating pixels $i$ and $j$.
The likelihood is a function of the parameters $\Teta$,
which may be either the $C_l$ or the world--model
constants, such as $\Omega$, etc...  In the latter
situation, the parameter dependence
enters the likelihood function through relations of the 
kind $C_l[\Teta]$, specified by the adopted theory, e.g.,
inflation.  In either case, the best estimates
for the parameter values are found by maximizing 
the likelihood function; confidence intervals can be
defined by treating the likelihood function as a 
probability distribution in $\Teta$ (with
non--uniform prior, if desired).

	We have implicitly been working with
simple pixel values, but it should be emphasized that all
follows through for temperature {\em differences}, or
any linear combination of sky temperatures, for these
simply transform the covariance matrix $\mC$.
Only Eq. (3) is altered; and it should
be noted that in the more general case, $T_{ij}$
may not be expandable in Legendre polynomials 
because, e.g., a difference measurement breaks spherical
symmetry (i.e., $\mT$ may depend on the 
orientation of the two difference pairs in a single 
difference scheme).  

	Experimental results are usually given in
terms of {\em band--powers}, estimates of the
variance of the temperature fluctuations over a finite 
range of $l$.  These may either
be defined by the differencing scheme employed
during observation, or by applying a linear
transformation to the pixel values of a map. 
For an ideal experiment with full sky coverage,
the band--powers would simply be the individual
$C_l$; however, limited sky coverage results
in less resolution in $l$--space, permitting
estimates only within finite bandwidths. 
A useful example is the single difference scheme, where
one measures $\Delta \equiv d_1 - d_2$ with 
$d_1$ and $d_2$ separated by an angle $\theta$ on
the sky.  In 
this case, the diagonal elements of the covariance
matrix may be written as
\begin{equation}
<\Delta^2>_{ens} = \frac{1}{4\pi} \sum_l (2l+1) C_l W_l
\end{equation}
where the {\em window function}, $W_l = 2|B_l|^2[1-P_l(\cos\theta)]$,
identifies the range of $l$ to which the the difference is 
sensitive.  The common approach is to quote a {\em flat band--power}
estimate, $\delta T_f$, defined by $(\delta T_f)^2 \equiv 
l(l+1)C_l/(2\pi)$, which leads to 
\begin{equation}
<\Delta^2>_{ens} = \frac{\delta T_f^2}{2} \sum_l  
\frac{(2l+1)}{l(l+1)} W_l
\end{equation}  
The band--power, $\delta T_f$, is then treated as the parameter
to be estimated from the full likelihood function.  It is this
procedure which leads to the points and uncertainties
shown in Figure 1.  We see that it
has indeed been constructed under the assumption of
gaussian fluctuations, as mentioned in the Introduction.
Notice also that because it contains all relevant 
information, the likelihood function includes
the uncertainty on the power estimate
due to sample, or
so--called ``cosmic'', variance.

	It is convenient, and perhaps
essential for future experiments, to 
use band--power estimates as the
starting point for constraining
cosmological parameters, instead of 
pixel vectors, as in Eq. (1).  Besides 
being the principle result reported in 
the literature, and hence easy to find, 
band--powers represent
a sort of data compression (Bond et al. 1998) --
there are fewer band--powers than pixels
for any given experiment, and hence fewer
calculations required to explore a given
parameter space.  If the fluctuations 
are truly gaussian, then we have lost nothing
in the compression.  To work in this direction,
we need to develop an easy--to--use approximation
to the full likelihood function for each
band--power estimate, ${\cal L}(\delta T_f)$,
one which hopefully requires little information
about experimental details.  

	With this aim, note first that the band--powers 
shown in Figure 1 are proportional to the variance 
of measured temperatures (or differences), e.g., as 
in Eq. (5).  To motivate an ansatz, consider
a totally unrealistic case where
the covariance matrix is strictly diagonal, 
including noise, and the noise is 
uniform with variance $\sig_N^2$:
\begin{equation}
C_{ij} = \left[(\delta T)^2 + \sig_N^2\right]\delta_{ij}
\end{equation}
Here, $(\delta T)^2$ represents the {\em true sky} 
variance (proportional to the band--power) we are trying to 
estimate.  In this case, we know that the maximum 
likelihood estimate of the variance of $\vd$,
including the noise, is 
$(1/\Npix)\sum_i^{\Npix} d_i^2$, i.e., this is the
best estimate of $(\delta T)^2 + \sig_N^2$.
We also know that the quantity 
$\chi_{\Npix}^2 \equiv (1/\Npix)\sum_i^{\Npix} d_i^2/
[(\delta T)^2 + \sig_N^2]$
is $\chi^2$--distributed with $\Npix$ 
degrees--of--freedom.  We may therefore express 
the best estimate of the sky variance as
\begin{equation}
[\delta T^{(e)}]^2 = [(\delta T)^2 + \sig_N^2] \chi_{\Npix}^2
	- \sig_N^2 
\end{equation}
and we know that {\em its} distribution, Prob$(\delta T^{(e)}|\delta T)$,
(in the frequentist sense) is related to a $\chi^2$--distribution
by the change of variable in Eq. (7).  Given an {\em actual} best estimate
of the sky variance from a particular experiment, $\delta T^{(o)}$, 
we argue that the likelihood function for $\delta T$ is 
${\cal L}(\delta T) = {\rm Prob}(\delta T^{(o)}|\delta T)$.

\begin{figure}
\centerline{\psfig{file=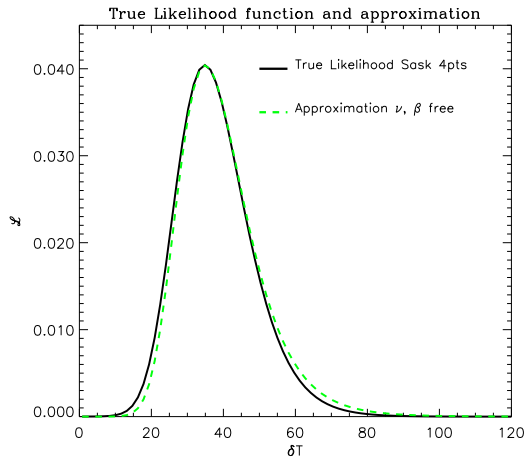, width=7cm}}
\caption{FIGURE 2. Comparison between the true likelihood function
for the 1995 Saskatoon Q--band 4--point difference, solid
line, with the approximation, shown as the dashed line (in green). 
The $x$--axis is in units of $\mu$K.  
}
\end{figure}


	These arguments apply only to this particular, simplistic
case.  For more general situations corresponding to 
actual experiments, we proceed by adopting the same functional form, 
${\cal L}(\delta T;\Npix,\sig_N^2)$, as an ansatz for the 
true likelihood function.  Notice that we now explicitly write
the dependence on both $\Npix$ and the noise.  What are these
quantities in real situations?  In some sense, they are
the number of independent pixels and an average noise level,
respectively.  Thus, using the number of pixels and the
given band--power estimate, we fit our approximation
by ajusting $\sig_N$ so that 68\% percent of the likelihood
falls within the quoted ($1\sigma$) error bars.  

	How well does this work?  The only way to answer that
question is by comparing the approximate likelihood 
to the true, complete likelihood function for a number of
experiments.  We have performed such a comparison for the
COBE, Saskatoon and MAX data sets.  In all cases, the approximation
works astoundingly well.  As an example, Figure 2 shows 
the comparison for a particular combination of the
Saskatoon data.

\begin{figure}
\centerline{\psfig{file=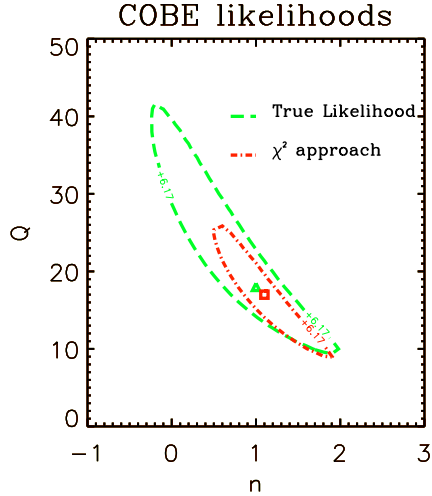, width=7cm}}
\caption{FIGURE 3. COBE likelihood contours on $Q$ and $n$ from the
true likelihood function, dashed lines (green), compared
to those from a $\chi^2$--analysis, dot--dashed lines
(inner contour in red).  The $Q$--axis is expressed in $\mu$K. 
}
\end{figure}


\bsk
\ni 3. CONCLUSIONS 
\ssk
\ni 
	Band--power estimates are not gaussian distributed,
and so use of the $\chi^2$--statistic to constrain models in the
power spectrum plane is not justified.  We have proposed 
an apparently good approximation to the likelihood function
which may be used as the basis of a more correct statistical
approach to the data in Figure 1.  Some preliminary details
of its application may be found in Bartlett et al. (1998b).
A general result seems to be that the constraints imposed by 
a $\chi^2$--analysis tend to be too strong (small confidence
regions) compared to the constraints from a more complete 
likelihood analysis.  An example of this in the case of COBE
is shown in Figure 3.

\bsk
\baselineskip = 12pt
{\abstract \ni ACKNOWLEDGMENTS

}

\bsk
\baselineskip = 12pt


{\references \ni REFERENCES
\ssk

\ref Bartlett, J.G., Blanchard, A., Le Dour, M., Douspis, M.
\& Barbosa, D. 1998a, in Fundamental Parameters in Cosmology,
Moriond 1998 proceedings, in press (astro--ph/9804158)

\ref Bartlett, J.G., Blanchard, A., Douspis, M. \& Le Dour, M.
1998b, in Evolution of Large--scale Structure: from Recombination
to Garching, MPA/ESO conference proceedings, in press

\ref Bond, J.R. \& Jaffe, A.H. 1996, in Microwave Background
Anisotropies (Moriond proceedings), F.R. Bouchet \&
R. Gispert (eds.), Editions Fronti\`eres, Gif--sur--Yvette, p.197

\ref Bond, J.R., Jaffe, A.H. \& Knox, L. 1998, astro--ph/9808264

\ref Borril, J. 1998, in 3K Cosmology, conference held at the
Universit\`a di Roma ``La Sapienza'', in press

\ref Hancock, S., Rocha, G., Lasenby, A.N. \& Gutierrez, C.M. 1998, 
MNRAS 296, L1

\ref Lineweaver, C., Barbosa, D., Blanchard, A. \& Bartlett, J.G.
1997, A\&A 322, 365

\ref Wandelt, B.D., Hivon, E. \& G\'orski, K.M. 1998, astro--ph/9808292
}                      

\end{document}